\documentclass{emulateapj}
\newcommand{\uv}{\mbox{$u$-$v$}}
\newcommand{\ex}[1]{\mbox{$\times 10^{#1}$}}

\newcommand{\kms}{\mbox{km s$^{-1}$}}
\newcommand{\Jb}{\mbox{Jy bm$^{-1}$}}
\newcommand{\muJb}{\mbox{$\mu$Jy bm$^{-1}$}}
\newcommand{\Ra}[4]{\mbox{${#1}^{\rm h} \; {#2}^{\rm m} \; {#3}\fs{#4} $}}
\newcommand{\dec}[4]{\mbox{${#1}\arcdeg \; {#2}\arcmin \; {#3}\farcs{#4} $}}
\newcommand{\thout}{\mbox{$\theta_{\rm o}$}}

\newcommand{\subNR}[1]{${#1}_{\rm NR}$}

\shortauthors{Bietenholz et al}
\shorttitle{VLBI Observations of SN 2009\lowercase{bb}}

\begin{document}
      
\title{VLBI Observations of the Type I b/c Supernova 2009\lowercase{bb}}

\author{M. F. Bietenholz\altaffilmark{1,2}, 
A. M. Soderberg\altaffilmark{3,4},
N. Bartel\altaffilmark{2},
S. P. Ellingsen\altaffilmark{5}, 
S. Horiuchi\altaffilmark{6},
C. J. Phillips\altaffilmark{7},
A. K. Tzioumis\altaffilmark{7},
M. H. Wieringa\altaffilmark{7}}
\author{N. N. Chugai\altaffilmark{8}}

\altaffiltext{1}{Hartebeesthoek Radio Observatory, PO Box 443, Krugersdorp, 
1740, South Africa} 
\altaffiltext{2}{Dept.\ of Physics and Astronomy, York University, Toronto,
M3J~1P3, Ontario, Canada}
\altaffiltext{3}{Harvard-Smithsonian Center for Astrophysics, Theory Division, 60
Garden Street, Cambridge, MA 02138, US}
\altaffiltext{4}{Hubble Fellow}
\altaffiltext{5}{School of Mathematics and Physics, University of Tasmania, Hobart, Tasmania, Australia}
\altaffiltext{6}{Canberra Deep Space Communication Complex, P.O. Box 1035, Tuggeranong, ACT 2901, Australia}
 
\altaffiltext{7}{Australia Telescope National Facility, Epping NSW, Australia}
\altaffiltext{8}{Institute of Astronomy, RAS, Pyatnitskaya 48, Moscow
119017, Russia}


\begin{abstract}
We report on VLBI, as well as VLA radio observations of the Type~Ib/c
supernova 2009bb.  The high radio luminosity of this supernova seems
to require relativistic outflow, implying that the early radio
emission was ``engine-driven'', that is driven by collimated outflow
from a compact object, even though no gamma-ray emission was seen.
The radio light curve shows a general decline, with a ``bump'' near $t
= 52$~d, seen most prominently at 5~GHz.  The lightcurve bump could be
either engine-driven, or it might represent the turn-on of the normal
radio emission from a supernova, driven by interaction with the CSM
rather than by the engine.
We undertook VLBI observations to resolve SN~2009bb's relativistic
outflow.  Our observations constrain the angular outer radius at an
age of 85~d to be $<0.64$~mas, corresponding to $<4\ex{17}$~cm and an
average apparent expansion speed of $<1.74\,c$.  This result is
consistent with the moderately relativistic ejecta speeds implied by
the radio luminosity and spectrum.
\end{abstract}


\section{INTRODUCTION}

Supernova \objectname{SN 2009bb} was discovered by the Chilean
Automatic Supernova Search Program \citep[CHASE;][]{Pignata+2009a,
Pignata+2009b} on 2009 March 29.9 UT, in the nearby spiral
galaxy \objectname{NGC 3278}.  
The radial velocity of NGC~3278 is 2964~\kms\ \citep{Paturel+2003},
and in what follows we will use a round distance of 40~Mpc for the
galaxy and the supernova.  SN~2009bb is in a region of high
star-formation, approximately 4.2~kpc from the center of the galaxy in
projection \citep{Levesque+2010}.  \citet{Stritzinger+2009} obtained
an optical spectrum which showed no evidence for hydrogen, and thus
SN~2009bb was classified as type I~b/c.  The shock breakout date is
well constrained to be March $19\pm1$~UT \citep{SN2009bb_Nature}.

Radio emission was discovered using the NRAO\footnote{The National
Radio Astronomy Observatory is a facility of the National Science
Foundation operated under cooperative agreement by Associated
Universities, Inc.}
Very Large Array (VLA) on April 5.2~UT, at time\footnote{We use $t$ to
refer to time in the observer frame.} after shock breakout, $t$ of
17~d.
The 8.5-GHz flux density was $24.5 \pm 1.2$~mJy
\citep{SN2009bb_Nature}.  This flux density corresponds to a spectral
luminosity of $\sim$5\ex{28}~erg~s$^{-1}$~Hz$^{-1}$, which is larger
than that observed for any other SN I~b/c at a similar time after
shock breakout \citep[see][]{SN2009bb_Nature, Soderberg+2006b,
Berger+2003}.  Subsequent VLA observations confirmed the initially
high flux density and showed a power-law decay, with the flux density
at 8.5~GHz, $S_{\rm 8.5 \, GHz} \propto t^{-1.4}$
\citep{SN2009bb_Nature}.  Similar decay rates are seen for other Type
I~b/c SNe, but also for the nearest gamma-ray burst, GRB~980425\@.
These radio observations gave a position for SN~2009bb of
\Ra{10}{31}{33}{87} and \dec{-39}{57}{30}{1} with an uncertainty of
0.7\arcsec\ in each coordinate.  We give this position, and all others
in this paper using J2000 coordinates.

Radio emission in a supernova is generated by the shocks formed as the
ejecta interact with the circumstellar material (CSM).  Radio emission
therefore traces the fastest ejecta, unlike the optical emission,
which traces the massive but more slowly-moving bulk of the ejecta.
Strong radio emission consequently is a sign of particularly strong
interaction with the CSM, which can be due either to a particularly
dense CSM, such as is seen for the radio-luminous Type II SNe, or to
particularly strong shocks which are caused by relativistic ejecta,
whether those latter are collimated or not.

Type I~b/c SNe like SN~2009bb are of special interest because GRBs
have been shown to be associated with them
\citep[e.g.,][]{Galama+1998,Stanek+2003,Malesani+2004,Pian+2006,
Cobb+2010, Starling+2010}.  While the optical luminosities of Type
I~b/c SNe and those associated with GRBs overlap, the GRBs are
distinguished by having powerful non-thermal ``afterglow'' emission.
In the radio, the afterglow typically peaks a few days after the
explosion, and GRB radio luminosities are observed to be up to a
million times higher than those of ordinary Type I~b/c SNe
\citep{Soderberg+2006c}.  This bright emission is the observational
manifestation of the substantial energy coupled to relativistic
velocities in GRBs.
However, GRBs are rare events, and \citet{Soderberg+2006b} showed
that less than 3\% of all Type I~b/c SNe have similarly relativistic
outflows.  The presence of relativistic ejecta, therefore, makes a
supernova of particular interest.
The physical mechanism that distinguishes ordinary Type I~b/c SNe from
GRB-SNe remains unknown and detailed studies of relativistic Type
I~b/c SNe are therefore required to make progress.

In particular, \citet{SN2009bb_Nature} showed that SN~2009bb's high
radio luminosity requires a substantial relativistic outflow powered
by a ``central engine'', in other words a black hole or a neutron star
surrounded by an accretion disk which produces collimated outflow.
The radio spectrum as measured at the VLA and the Giant Meterwave
Radio Telescope (GMRT) is well fit by a synchrotron self-absorption
(SSA) spectrum.  The high luminosity and relatively low turnover
frequency of $\sim$6~GHz at $t = 20$~d then imply a blastwave radius
of $4.4$\ex{16}~cm \citep{SN2009bb_Nature} and therefore a mean
apparent expansion speed of $0.85 \pm 0.02 c$, assuming equipartition
of energy between electrons and magnetic fields.  Note that these are
minimum values for the size and expansion velocity, since both
deviations from equipartition, and the presense of free-free
absorption (FFA) in addition to SSA would result in larger values.  No
gamma-ray counterpart was detected, but an off-axis viewing angle or a
low fluence burst cannot be excluded.  SN~2009bb differs from GRBs in
that it occurred in a high-metallicity environment
\citep{Levesque+2010}.

For relatively nearby SNe, a direct measurement of the size of the
shockwave is possible with very-long-baseline interferometry (VLBI)
observations.  Such a measurement provides a model-independent way of
measuring the expansion speed and therefore determining the presence
or absence of relativistic ejecta, and possibly also determining the
emission geometry and testing the assumption of equipartition.
Unfortunately, SNe which are both sufficiently nearby and radio-bright
to allow VLBI imaging are rare events.  Relativistic expansion was
clearly detected using VLBI in the case of GRB~030329
\citep{Taylor+2004}.  In the case of SNe not associated with GRBs,
however, the VLBI observations so far have confirmed the rarity of
relativistic ejecta: VLBI observations of two Type~I~b/c SNe which
were suspected of having relativistic ejecta, SN~2008D and SN~2001em,
showed only subluminal expansion \citep{SN2008D-VLBI, Paragi+2008,
Schinzel+2008, SN2001em-2, SN2001em-1, Paragi+2005}.  In the case of
SN~2007gr, relativistic expansion was claimed by
\citet{SN2007gr_Nature}, but \citet{SN2007gr-Soderberg} showed that a
more conservative interpretation of a normal, non-relativistic
supernova can also be reconciled with the VLBI measurements, and
provides a more natural explanation of the relatively low radio and
X-ray luminosity.  Optical and infra-red spectra also suggest a modest
ejected mass and explosion energy \citep{Mazzali+2010}, whereas
relativistic ejecta are usually accompanied by large ejected masses
and explosion energies.

\section{OBSERVATIONS}

We obtained both VLA total-flux-density and VLBI imaging observations
of SN~2009bb.  The VLA observations were obtained in two ways.  Firstly,
by using the VLA as a standalone interferometer in parallel with the
use of the phased VLA as part of our VLBI array on 2009 June 10 - 11,
described below.  These observations were at 8.4~GHz in the CnB array
configuration. Secondly, as part of a regular VLA monitoring program
for Type I b/c supernovae (AS983; PI Soderberg)
on 2009 October 23, at both 8.4 and 5.0~GHz, and with the array in the
D configuration.  The observations were reduced in a standard manner,
with the flux density scale being set by observations of 3C~286, and
\objectname[]{VCS4 J1036-3744} (hereafter J1036-3744; also known as
QSO B1034-374) being used for phase-referencing.

The VLBI observations were carried out at 8.4~GHz, and lasted for 10
hours with a midpoint of 2009 June 12.1 UT, or $t = 85$~d.
Our VLBI array consisted of the NRAO VLBA ($10 \times 25$-m
diameter) and
phased VLA (130-m equivalent diameter) telescopes, and the Hobart
(25-m diameter) and Tidbinbilla DSS45 (34-m diameter) telescopes of
the Australian Long Baseline Array.  Unfortunately, due to a bearing
failure, the 26-m antenna at Hartebeesthoek, South Africa was not
available for these observations. We recorded a bandwidth of 64~MHz in
both senses of circular polarization with two-bit sampling, for a
total bit rate 512~Mbit~s$^{-1}$, with the exception of the
Tidbinbilla antenna, at which we only recorded left circular
polarization (IEEE convention).
The VLBI data were correlated with NRAO's VLBA processor, and the
analysis carried out with NRAO's Astronomical Image Processing System
(AIPS)\@.  The initial flux density calibration was done through
measurements of the system temperature at each telescope, and then
improved through selfcalibration of the reference source.  A
correction was made for the dispersive delay due to the ionosphere
using the AIPS task TECOR, although the effect at our frequency is not
large.

\begin{figure}
\centering
\includegraphics[width=0.46\textwidth]{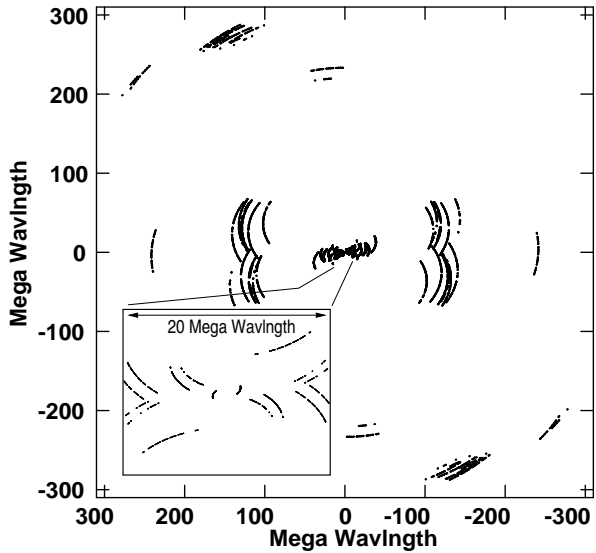}
\caption{The \uv-coverage obtained for our VLBI observating run for
SN~2009bb on 2009 June 10 - 11 at 8.4~GHz after editing and excluding
measurements with antenna elevations below 12\arcdeg.  The inset shows
a detail of the $\pm$10~M$\lambda \times \pm$8~M$\lambda$ central
region of the \uv~plane showing the coverage of the shortest
baselines.}
\label{fuvcov}
\end{figure}

We phase-referenced our VLBI observations to J1036-3744, for which we
use the position from the Fourth VLBA calibrator survey of
\Ra{10}{36}{53}{43961}, \dec{-37}{44}{15}{0662}
\citep{Petrov+2006}. We used a cycle time of $\sim$4~min, with
$\sim$2.7~min spent on SN~2009bb.  Due to SN~2009bb's southern
declination, it was at relatively low elevation at the North American
antennas. The source never exceeded 8\arcdeg\ in elevation at the BR,
HN and NL antennas of the VLBA, therefore we did not use the data from
these antennas in the final analysis.  The majority of our remaining
visibility measurements were made at relatively low elevations, with
less than 25\% of our individual visibility measurements being
obtained with both antennas observing at elevations $>15\arcdeg$.  We
show the final \uv-coverage obtained in Figure~\ref{fuvcov}.

For both imaging and model-fitting of the VLBI data for SN~2009bb, we
reduced the weights of the VLA by a factor of 4\@. As the VLA is more
than an order of magnitude more sensitive than any of the remainder of the
antennas, the fraction of visibilities involving the VLA will have
much higher weight and thus dominate.  Although reducing the VLA
weights incurs a penalty in statistical efficiency, it improves the
stability of both imaging and modelfitting.

\section{RESULTS}

\subsection{VLA Total Flux Density}
\label{svlaflux}

We describe first the results from the reduction of the 8.4~GHz VLA
interferometric data.  Our flux-density uncertainties include both
statistical standard errors and an assumed 5\% uncertainty in the flux
density scale, added in quadrature.  For our calibrator source,
J1036-3744, we obtained a flux density of
$0.77 \pm 0.04$~Jy.  

\begin{figure}
\centering
\includegraphics[width=0.46\textwidth]{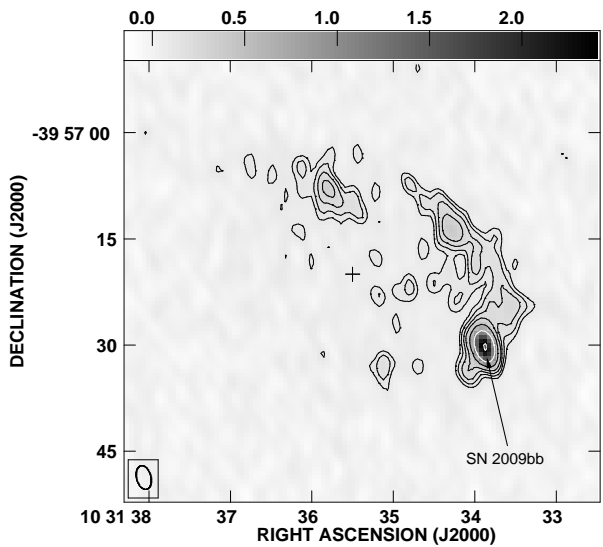}
\caption{An 8.4-GHz VLA image of NGC~3278 on 2009 June 12\@.
SN~2009bb is indicated.  The contours are at $-0.1$, 0.1, 0.14, 0.20,
0.28, 0.5, 1.0, and 2.5 m\Jb, and the greyscale is labeled in m\Jb.
The peak brightness in this sub-image was 2640~$\mu$\Jb, the
background rms 31~$\mu$\Jb, and the FWHM of the convolving beam,
indicated at lower left, was $3.42\arcsec \times 2.03\arcsec$ at p.a.\
14\arcdeg.  The position of the galaxy center \citep{Paturel+2003} is
indicated by a cross.}
\label{fngc}
\end{figure}

We show the VLA image of NGC~3278 and SN~2009bb in Figure~\ref{fngc}.
The presence of a small amount of extended emission due to the galaxy
NGC~3278 is apparent.

By fitting an elliptical Gaussian of the same dimensions as the
restoring beam and a zero level, we determine the 8.4-GHz flux density
of SN~2009bb at $t = 85$~d to be $2.47 \pm 0.19$~mJy, where our
uncertainty includes a contribution due to the uncertainty in
estimating the zero-level (as well as the aforementioned 5\%
uncertainty in the flux-density scale, all added in quadrature).  We
estimate the contribution from NGC~3278 to be $240 \pm 100 \, \mu$\Jb\
at our resolution of $3.42\arcsec \times 2.03\arcsec$ (FWHM).

For our subsequent VLA observations on 2009 Oct.\ 23, the array was in
the lowest-resolution D array configuration with a synthesized
beamwidth of $31\arcsec \times 6\arcsec$ at 8.4~GHz and $47\arcsec\
\times 10\arcsec$ at 5.0~GHz. The subtraction of the galaxy component
of the radio emission was not straightforward.  We accomplished it by
using the image of SN~2009bb and NGC~3278 made on 12 June 2009
(Figure~\ref{fngc}), which had adequate resolution and good
\uv~coverage, as a template.  We assume that NGC~3278 does not change
with time, and that any change in the image between June and October
is therefore due to SN~2009bb.  We convolved the template image to the
resolution of the new ones, and in the case of 5~GHz, scaled the
brightness distribution by an assumed spectral index of $-0.6$, and
then determined the change in the flux density of SN~2009bb.  We find
the flux density of SN~2009bb on 2009 Oct.\ 23 ($t = 218$~d) was $0.97
\pm 0.24$~mJy at 8.4~GHz and $1.9 \pm 1.0$~mJy at 5.0~GHz.  Our
uncertainties again include a contribution from the galaxy
subtraction, as well as those from noise and the flux-density
calibration, all added in quadrature.

\begin{figure}
\centering
\includegraphics[width=0.46\textwidth]{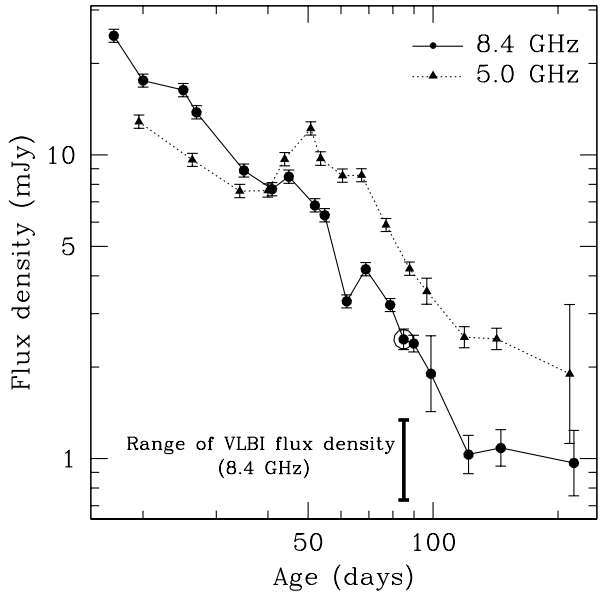}
\caption{The 4.8 and 8.4-GHz radio light curve of SN~2009bb from VLA
measurements.  The measurements are described in this paper or in
\citet{SN2009bb_Nature}.  Our uncertainties include an estimated 5\%
uncertainty in the VLA flux-density scale and a contribution from the
galaxy subtraction.  The VLA flux density measurement corresponding to
our VLBI observations is circled.  For better visibility, we have
shifted the 5-GHz points very slightly in time so as to avoid
overlapping error bars.  We also indicate the range of flux densities
recovered from the VLBI observations to illustrate the discrepancy
between them and the VLA flux densities (see appendix for details).}
\label{flightcurve}
\end{figure}

In Figure \ref{flightcurve}, we plot the radio light curves of
SN~2009bb at 8.4 and 5~GHz including the flux density measurements
described above, as well as earlier values reported in
\citet{SN2009bb_Nature}.  Note that the VLA flux density obtained for
the date and frequency of the VLBI observations ($t = 85$~d, 8.4~GHz)
shows no noticeable discrepancy with the remainder of the observed
light curve.

A bump in the light curves is observed near $t \simeq 52$~d.  It is
most prominent in the 5-GHz light curve, which shows an increase by
almost a factor of 2, reaching a local maximum of 12~mJy at $t \simeq
52$~d, which corresponds to a spectral luminosity of
2.3\ex{28}~erg~s$^{-1}$~Hz$^{-1}$.  A similar although smaller
relative increase is seen at 8.4~GHz, with the peak occurring perhaps
slightly earlier.

A weighted least-squares fit to all the data up to $t = 220$~d shows
that on average, the flux density decays as follows: $S_{\rm 8.5 \,
GHz} \propto t^{-1.4 \pm 0.1}$ and $S_{\rm 5 \, GHz} \propto t^{-0.7
\pm 0.2}$.  There is a possible flattening of the flux density decay
at the very latest times ($t > 100$~d), although due to the difficulty
of galaxy subtraction the reality of any late-time flattening is
difficult to confirm.  While the overall average decay rate is
significantly flatter at 5~GHz than at 8.4~GHz, this is largely due to
the bump being more prominent at 5~GHz, with the decay rates being
similar at both frequencies after the bump.

\subsection{VLBI}
\label{svlbi}

We now discuss the results of the VLBI observations, turning first to
our phase-calibrator source, J1036-3744, as the results obtained for
it will inform our interpretation of the SN~2009bb results.  From a
deconvolved (CLEAN) image of J1036-3744, we recovered a total flux
density of 687~mJy, with an background rms brightness of 1.6~m\Jb. The
image is dynamic-range limited, rather than limited by thermal noise.
The total flux density recovered is 10\% less than the flux density
measured at the VLA\@.
It is possible this discrepancy is due to 10\% of the flux density
being at spatial scales too large for VLBI but too small for the VLA
($0.03\arcsec \sim 0.3\arcsec$), although it is not uncommon to see
such discrepancies between the flux density scales for VLBI,
determined from the system-temperature measurements, from the more
accurate one for VLA measurements, determined by observations of
calibrator sources such as 3C~286\@.
The residual delays and delay-rates we found for J1036-3744 were
moderate, being mostly $<40$~nsec and $<5$~milli-Hz respectively,
suggesting that the array was performing well, and therefore no
particularly strong decorrelation is expected.

We turn now to SN~2009bb.  An image made using complex weighting with
the robustness factor set to $-2$, which is close to uniform
weighting, and a \uv~taper of 30\% at 120~M$\lambda$ had a total clean
flux density of 740~$\mu$Jy, a peak brightness of 610~$\mu$\Jb, and a
background rms brightness of 130~$\mu$\Jb.
We tried different weighting schemes and \uv~tapers, but in no case
was the total clean flux density greater than 800~$\mu$Jy.

For marginally resolved sources, such as SN~2009bb, the best values
for the source size and VLBI flux density come from fitting models
directly to the visibility data, rather than imaging.  We choose as a
model the projection of an optically-thin spherical shell of uniform
volume emissivity, with an outer radius of $1.25\times$ the inner
one\footnote{Our results do not depend significantly on the assumed
shell thickness, as the effect of reasonable thicknesses different
than the assumed one is considerably less than our stated
uncertainties.}.
Such a model has been found to be appropriate for other radio
supernovae \citep[see e.g.,][]{SN93J-3, SN79C-shell}.  For a partially
resolved source such as SN~2009bb, the exact model geometry is not
critical, and our shell model will give a reasonable estimate of the
size of any circularly symmetric source, with a scaling factor of
order unity dependent on the exact morphology \citep[see discussion
in][]{SN93J-2}.  The Fourier transform of this shell model is then fit
to the visibility measurements by least squares.

For the interested reader, we give the details of the modelfitting
results in the appendix.  Fitting such a model to the strictly
phase-referenced visibilities for SN~2009bb gives our most accurate
estimate of its center position, which is
\Ra{10}{31}{33}{8762}, \dec{-39}{57}{30}{022}, with an estimated
uncertainty, dominated by systematic contributions, of a few mas.

The total flux density recovered from the VLBI data, whether through
imaging or modelfitting, was at most $\sim$50\% of that measured at
the VLA\@.  This discrepancy might suggest that the source is
over-resolved by the VLBI observations, or, more precisely, that the
missing flux density is at angular scales too large to be seen in the
VLBI observations, but smaller than the VLA resolution.  The VLBI
observations had reasonable \uv~coverage even for baselines as short
as 10~M$\lambda$ (see inset in Figure~\ref{fuvcov}), and are therefore
sensitive to structure up to $\sim$20~mas in angular size, while the
VLA resolution was $\sim$3\arcsec.  If the discrepancy between the VLA
and VLBI flux densities is to be ascribed to a source resolved in
VLBI, then its size should therefore be in the range of 20~mas to
3\arcsec.  The minimum angular size makes it unlikely that such
hypothetical structure could be related to SN~2009bb, since for a
supernova age of 85~d and the distance 40~Mpc, the implied apparent
expansion speed is $> 25 \, c$, which we consider improbable.

Another possibility is a source of radio emission unrelated to
SN~2009bb but coincidentally so close as to be within the VLA
beamwidth.  We consider also this possibility to be unlikely. A flux
density of at least 1~mJy is required to explain the discrepancy,
while the brightest of NGC~3278's galactic radio emission seen at the
VLA was only $\sim$400~\muJb, and appears well-resolved
(Figure~\ref{fngc}).  It would seem improbable therefore that there
would be such a bright and compact unrelated source of emission
located so close to SN~2009bb.

The third possibility, which we consider most likely, is that there is
substantial decorrelation in the VLBI measurements, due to uncorrected
differences in atmospheric delay between SN~2009bb and the
phase-reference source\footnote{We note that inaccuracies in the
correlator model could also cause coherence loss, but are typically
considerably smaller than the un-modelled contributions of the
atmosphere.}.
In other words, the phase-calibration for SN~2009bb is of poor
quality.  Any determination of the source size must therefore be
interpreted with caution.  For a sufficiently strong source, the phase
calibration could be improved by selfcalibration.  SN~2009bb, however,
was not bright enough to allow conventional selfcalibration using an
image.  As an alternative, we introduced the antenna phases as free
parameters in the \uv~plane modelfit.  Unfortunately, due to the poor
\uv~coverage and the low signal-to-noise ratio, we were only able to
obtain an upper limit of 0.64~mas to the source radius with this
procedure.  We found, however, that even with this effective phase
selfcalibration, we were not able to recover more than about half of
the VLA flux density.  Our upper limit on SN~2009bb's angular size is
not a formal statistical limit because of various assumptions made in
its derivation, but it is intended as a $3\sigma$ limit, and should be
reasonably robust.  We describe the exact procedure by which we arrive
at this limit in the appendix for the interested reader.

Since the age of SN~2009bb at the time of our VLBI observations was
85~d, using a distance of 40~Mpc as well as assuming circular
symmetry, this limit on the angular radius allows us to obtain a
corresponding upper limit on the apparent expansion speed of
$1.74\,c$.

\section{DISCUSSION}

We have made VLA and VLBI observations of the type I b/c supernova
2009bb. This supernova was of particular interest because the high
level of radio emission showed that substantial material was ejected
at relativistic speeds \citep{SN2009bb_Nature}.  We discuss first the
evolution of the radio light curve as revealed by our VLA measurements
of the total flux density

\subsection{Radio Light Curve Evolution}

\begin{deluxetable}{l r r r r }
\tabletypesize{\small}
\tablecaption{Size Estimates for SN 2009\protect\lowercase{bb}}
\tablehead{
\colhead{Age (days)\tablenotemark{a}} &
\colhead{20} & \colhead{52} & \colhead{81} & \colhead{145}
}
\startdata
Radius\tablenotemark{b} ($\times 10^{16}$ cm) &  4.4 &  8.0 &  22 &  26 \\
Angular Radius (mas)\tablenotemark{c}         & 0.073& 0.13 & 0.36 & 0.43 \\
Average Apparent Velocity 
                                              & 0.85 & 0.60 & 0.71 & 0.69 
\enddata
\tablenotetext{a}{Age in days since shock breakout.}
\tablenotetext{b}{Linear size computed from radio spectrum by assuming SSA to be
the dominant absorption mechanism. Radio spectra in \citet{SN2009bb_Nature}.}
\tablenotetext{c}{Angular radius (for $D = 40$~Mpc).}
\label{tsize}
\end{deluxetable}

Our new data reveal a continuing decline of the radio light curve of
SN~2009bb at late times (Figure \ref{flightcurve}).
By assuming that the turnover in the radio spectrum is due to SSA, we
can estimate the size at different epochs from the four radio spectra
published in \citet{SN2009bb_Nature}.  We give these values of the
radius, along with the corresponding values angular radius and the
average expansion velocity as a fraction of $c$ in Table \ref{tsize}.
These are minimum radii: if the spectral turnover were due to, for
example, FFA rather than SSA, the supernova would be larger.  However,
our VLBI measurement gives a $3\sigma$ upper limit on the radius of
0.64~mas at $t = 85$~d, which rules out a radius much larger than
those derived from SSA\@.  In the case of SN~2009bb, much larger sizes
and expansion velocities are probably ruled out also on energetic
grounds, as \citet{SN2009bb_Nature} showed that the above size
estimates imply an energy of $(1.3 \pm 0.1) \times 10^{49}$ erg
coupled to relativstic ejecta, and much larger amounts of energy
coupled to the relativistic ejecta are improbable.  More generally,
\citet{Chevalier1998} shows that SSA is probably the dominant
absorption mechanism for Type I b/c SNe. In summary, we think it
unlikely that the sizes or velocities in Table~\ref{tsize} are
substantially in error.
These radius values suggests deceleration
between $t$ = 20~d and 52~d, but possibly a re-acceleration between
$t$ = 50~d and 81~d, and a relatively constant speed expansion since
then.

A bump is visible in the light curves, which is most
prominent at 5~GHz, where it peaks at $t \simeq 52$~d, and the flux
density increases by a factor of $\sim$2 relative to the longer-term
decay. In terms of spectral luminosity, the bump represents a relative
increase of $\sim$1.2\ex{28}~erg~s$^{-1}$~Hz$^{-1}$.

What is the origin of this bump?  As mentioned, SN~2009bb was
distinguished by having mildly relativistic ejecta.
One hypothesis is that the bump could be driven by the ``engine'', in
other words represent a renewed injection of energy: similar bumps in
the lightcurve were seen for GRB~980425/SN~1998bw
\citep{Kulkarni+1998, LiC1999} during the relativistic phase, and are
thought to be engine-driven.  The fact that the SSA sizes derived
above suggest a re-acceleration between $t$=50~d and 82~d would be
consistent with this interpretation, in other words that there was a
renewed energy input at the forward shock somewhere around $t = 50$~d.

Alternatively, the bump may be related to the collision between the
bulk of the ejecta, which are almost undecelerated, with the leading
blast wave, which was initially relativistic but decelerates as it
sweeps up the stellar wind.  It is not straightforward to predict the
outcome of this collision because it essentially depends on the poorly
known distributions of density and velocity in the mildly
relativistic, outermost layers of SN ejecta. We note, however, that
this scenario is similar in many respects to that of the bump
representing renewed energy input by the engine: in both cases the
bump in the light curve is the result of the collision between the as
yet undecelerated inner shell and the decelerating external shock. The
difference may be in a possible deviation from a spherical symmetry,
which is expected to be larger in the scenario of recurrent jets than
in the present case. To summarize, we consider the bump in the light
curve likely to be engine-driven, although the collision of the main
ejecta with the external shock wave is not excluded.

Interestingly, even the case that bump is engine-driven, i.e., due to
recurrent jet activity, the collision of the main SN ejecta with the
decelerating blast wave is unavoidable.  Such a collision might
therefore be expected in supernovae similar to SN~2009bb or SN~1998bw
at $t = 50 \sim 100$~d, and may produce detectable effects in the
radio lightcurves.


\subsection{Source Size Derived from VLBI Observations}

The primary goal of our VLBI observations was to set limits on the
expansion speed by measuring the angular size of SN~2009bb.
Unfortunately, the southern declination of the source meant that it
was at low elevation for most of our VLBI array, which limited the
quality of the data obtained, and consequently the accuracy of the
determination of the source's angular size.

We obtained an estimate of the source outer angular radius between 0
and 0.64~mas, suggesting average apparent expansion speeds between 0
and $1.74c$, with the limits intended to represent a $3\sigma$ range
(see \S~\ref{svlbi} and the appendix for details of our size
determination and the caveats applying thereto).  This range of
apparent expansion speeds implies a radius at $t = 85$~d of $<
4\ex{17}$~cm and a bulk Lorentz factor of $< 2.0$.

\citet{SN2009bb_Nature} showed that deceleration had likely occurred
relatively early in SN~2009bb, with time of the non-relativistic
transition \subNR{t} $\simeq 1$~d, occurring well before our VLBI
observations.  The recent simulations of aspherical supernova
explosions in stripped-envelope stars by \citet{Couch+2010} and
\citet{Kifonidis+2006} have also shown that even very aspherical
energy releases have a largely spherical outer envelope by $t \simeq
1$~d.
We would therefore expect an approximately spherical blast-wave at the
time of our observations.  

\citet{SN2009bb_Nature} also determined that at $t = 20$~d, the radius
of the supernova was $\sim$4.4\ex{16}~cm. If we assume a standard
Sedov-von Neumann-Taylor evolution \citep[see e.g.,][]{GranotL2003}
after $t = 20$~d, we would expect a radius at the time of our VLBI
observations ($t = 85$~d) of $\sim$8\ex{16}~cm which is well within
our observational limits on radius.

VLBI observations of SNe such as SN~2009bb are crucial to directly
measuring the expansion speeds and perhaps the geometry of the radio
emission, and thus important to confirming a jet model.  However, such
VLBI observations are difficult and generally limited by the available
sensitivity and \uv~coverage, especially if the supernova is at a
southern declination.  They should become easier in the future with
the planned increases in sensitivity of the VLBA
\citep{Ulvestad+2010}, and by future availability of South African and
Australian SKA pathfinder instruments, MeerKAT \citep{Booth+2009} and
ASKAP \citep{Johnston+2008} respectively, as VLBI stations.

\section*{APPENDIX}

In this appendix, we describe in detail how we arrived at the upper
limit on the angular size of SN~2009bb from the VLBI observations.  We
first made an image from the strictly phase-referenced VLBI data for
SN~2009bb, using robust weighting and dropping any data taken with
telescope elevations below 12\arcdeg.  We found a peak brightness of
only $610 \; \mu$\Jb\ for a convolving beam of $3.4 \times 0.8$~mas at
p.a.\ 0\arcdeg\ (FWHM).  Even if we restrict ourselves to the
baseline-lengths of $<10$~M$\lambda$,we found a peak brightness of
only $800 \pm 160 \; \mu$\Jb, with a convolving beam size of $42 \times
10$~mas.
We note however, that due to the uneven \uv~coverage, the sidelobe
levels are quite high, reaching peaks $>70$\%, and that deconvolution
is therefore likely not very reliable.  

As an alternative to imaging, avoiding the need for deconvolution, we
fit a spherical shell model directly to the complex visibilities by
least squares as described in \S~\ref{svlbi} above.  When fitting the
strictly phase-referenced visibility data, the best-fit model had a
total flux density of $730 \pm 80 \, \mu$Jy with an outer angular radius,
\thout, of 0.47~mas, and a best-fit center position of
\Ra{10}{31}{33}{8762}, \dec{-39}{57}{30}{022}.  This is the position
estimate given in \S~\ref{svlbi} above.

The total flux density of 730~$\mu$Jy recovered from strictly
phase-referenced data is considerably below the value of $2.47 \pm
0.19$~mJy measured at the VLA (\S~\ref{svlaflux}).  As detailed in
\S~\ref{svlbi} we think this discrepancy is unlikely to be due to
either a supernova which is so large as to be over-resolved by the VLBI
measurements or to an unrelated nearby source of emission.  The
discrepancy is therefore most likely due to decorrelation, in other
words poor phase calibration. Since most of observations were of
necessity made at relativly low elevations, difficulties
in calibration are perhaps not unexpected.

Our phase-calibration for J1036-3744 is of good quality, as attested
to by our recovery of 90\% of its VLA flux density.  The delay changes
relatively smoothly as a function of time, so there is no reason to
suspect the interpolation in time between the J1036-374 scans to the
intervening SN~2009bb scans.  SN~2009bb, however, is 2.5\arcdeg\ away
on the sky, with the difference being predominately in declination.
Any un-modeled elevation dependence of the delay will therefore result
in errors in the delay for SN~2009bb, and poor phase-referencing.  The
sitution for SN~2009bb is particularly bad since the difference in
source position is mostly in declination and since the observations
are mostly at low elevation where the airmass is large.
Self calibration in phase would in principle allow improving the
calibration of the supernova visibilities, although it does have the
drawback of introducing biases \citep[see e.g.,][]{MassiA1999,
Marti-VidalM2008}, which can be severe in the case of low
signal-to-noise.

As our signal-to-noise ratio is too low for traditional
selfcalibration using images, we instead introduce the phases of the
complex antenna gains as free parameters in the model-fitting
procedure, using a slightly modified version of the AIPS task OMFIT\@.
This procedure has the advantage of allowing a more quantitative
measure of the goodness-of-fit than traditional selfcalibration using
images.  Due to the low signal-to-noise, we fitted for only a single
phase solution common all 8 intermediate-frequency channels.  We fix
the source position at the best-fit position obtained above, and let
the antenna phases vary on a 30-min timescale.  The best fit model for
SN~2009bb obtained in this way had $\thout = 0.22^{+0.06}_{-0.08}$~mas
and a flux density of $1.34 \pm 0.07$~mJy (statistical uncertainties).
Even in this case, the total flux density in the model was only 54\%
of that measured at the VLA.

For a partly resolved source, the fitted source size is generally also
correlated with the antenna amplitude gains.  We tested for an
additional uncertainty due to mis-calibration of the antenna amplitude
gains by artificially varying individual antenna gains by $\pm 25\%$,
and then fitting a model to SN~2009bb as above.  The resulting rms
variation in \thout\ was 0.05 mas.  Adding this to the above
uncertainties in quadrature results in a value for \thout\ of
$0.22^{+0.08}_{-0.09}$~mas.

As mentioned, selfcalibration can introduce biases.  To test for this
possibility, we calculated simulated visibilities from models with
various values of \thout\ and random noise at a level corresponding to
that in our observations.  We then fit these simulated visibilities
using the same procedure as above, including the addition of the
antenna phases as free parameters.  These tests suggests that the true
value of \thout\ is in fact $\sim$18\% higher than that determined
from the fitting.  For simplicity, we carried out this test using a
disk, rather than a spherical shell model, that the relative bias in
\thout\ should be very similar.  In other words, selfcalibration tends
to make the source appear more compact.

Correcting for this bias, we thus arrive at a final, unbiased estimate
of \thout\ for SN~2009bb of $0.26^{+0.09}_{-0.11}$~mas, with a
$3\sigma$ upper limit of 0.59~mas. However as noted above, even with
phase-selfcalibration, the fitted flux density is still notably below
that measured by the VLA, which suggests the presence of further
decorrelation (or some other source of error).  As the signal-to-noise
is already lower than is generally considered safe for
selfcalibration, using a shorter solution interval is not advisable.

As a final test, we fixed the model flux density at 2.0~mJy, a round
value slightly below but near that measured by the VLA\@, and again
fitted a model of the source as well as the antenna gain-phases.  We
find that the best fit to the VLBI visibilities is not much larger
than the above estimates, having $\thout = 0.34_{-0.04}^{+0.10}$~mas,
with a statistical $3\sigma$ upper limit of 0.64~mas.  We note that forcing
such a large flux density on the model results in a significant
increase in $\chi^2$ over models with lower flux density.

The VLBI visibilities, therefore, seem to robustly suggest a total
flux density for SN~2009bb of $\lesssim 50$\% of that observed at the
VLA\@.  Given this inconsistency, the biases involved in
phase-selfcalibration, and the likelihood of coherence losses not
accounted for by our phase-selfcalibration, we suggest a probable
range for the outer radius of SN~2009bb of $0 < \thout < 0.64$~mas.
The failure to recover the total flux density suggests the possibility
that significant decorrelation remains in the VLBI data, but such
decorrelation is more likely to increase the apparent size of the
source than decrease it, so our upper limit on the angular size should
be robust.

A very similar phenomenon was seen in the case of SN~2007gr.  Also for
this SN, the flux density recovered from VLBI observations was
considerably lower than the total flux density measured by a
connected-element interferometer.  Initially, this discrepancy was
interpreted as suggesting a large source size and thus relativistic
expansion \citep{SN2007gr_Nature}.  This was somewhat surprising,
given that SN~2007gr's peak 8.4-GHz spectral luminosity was relatively
low, being $\sim$500 times lower than that of SN~2009bb.
\citet{SN2007gr-Soderberg} showed that the radio lightcurves and the
lack of detectable X-ray emission were fully consistent with a normal,
non-relativistic SN, but were in fact hard to reconcile with
relativistic expansion.  They also re-examined the SN~2007gr VLBI data
and showed that the low VLBI flux density was observed on both short
and long baselines, and if it was to be explained by a large,
heavily-resolved source, required very large apparent expansion
velocities of $>2c$.
They conclude that coherence losses which were larger than normal but
not improbably so provided an explanation which as plausible as the
original one of modestly relativistic expansion for SN~2007gr.  In the
present case of SN~2009bb, some loss of coherence is not unexpected,
given that the southern declination of the source necessitated
observations made mostly at low elevation.

\acknowledgements 
\noindent{Research at York University was partly supported by NSERC\@.
We have made use of NASA's Astrophysics Data System Bibliographic
Services.}

\bibliographystyle{hapj}
\bibliography{mybib1}

\begin{thebibliography}{40}
\expandafter\ifx\csname natexlab\endcsname\relax\def\natexlab#1{#1}\fi

\bibitem[{{Bartel} \& {Bietenholz}(2008)}]{SN79C-shell}
{Bartel}, N., \& {Bietenholz}, M.~F. 2008, \apj, 682, 1065, arXiv:0806.3482

\bibitem[{{Bartel} {et~al.}(2002){Bartel}, {Bietenholz}, {Rupen}, {Beasley},
  {Graham}, {Altunin}, {Venturi}, {Umana}, {Cannon}, \& {Conway}}]{SN93J-2}
{Bartel}, N. {et~al.} 2002, \apj, 581, 404

\bibitem[{{Berger} {et~al.}(2003){Berger}, {Kulkarni}, {Frail}, \&
  {Soderberg}}]{Berger+2003}
{Berger}, E., {Kulkarni}, S.~R., {Frail}, D.~A., \& {Soderberg}, A.~M. 2003,
  \apj, 599, 408, arXiv:astro-ph/0307228

\bibitem[{{Bietenholz} \& {Bartel}(2005)}]{SN2001em-1}
{Bietenholz}, M.~F., \& {Bartel}, N. 2005, \apjl, 625, L99

\bibitem[{{Bietenholz} \& {Bartel}(2007)}]{SN2001em-2}
------. 2007, \apjl, 665, L47, arXiv:0706.3344

\bibitem[{{Bietenholz} {et~al.}(2003){Bietenholz}, {Bartel}, \&
  {Rupen}}]{SN93J-3}
{Bietenholz}, M.~F., {Bartel}, N., \& {Rupen}, M.~P. 2003, \apj, 597, 374,
  arXiv:astro-ph/0307382

\bibitem[{{Bietenholz} {et~al.}(2009){Bietenholz}, {Soderberg}, \&
  {Bartel}}]{SN2008D-VLBI}
{Bietenholz}, M.~F., {Soderberg}, A.~M., \& {Bartel}, N. 2009, \apjl, 694, L6

\bibitem[{{Booth} {et~al.}(2009){Booth}, {de Blok}, {Jonas}, \&
  {Fanaroff}}]{Booth+2009}
{Booth}, R.~S., {de Blok}, W.~J.~G., {Jonas}, J.~L., \& {Fanaroff}, B. 2009,
  ArXiv e-prints, 0910.2935

\bibitem[{{Chevalier}(1998)}]{Chevalier1998}
{Chevalier}, R.~A. 1998, \apj, 499, 810

\bibitem[{{Cobb} {et~al.}(2010){Cobb}, {Bloom}, {Perley}, {Morgan}, {Cenko}, \&
  {Filippenko}}]{Cobb+2010}
{Cobb}, B.~E., {Bloom}, J.~S., {Perley}, D.~A., {Morgan}, A.~N., {Cenko},
  S.~B., \& {Filippenko}, A.~V. 2010, \apjl, 718, L150, 1005.4961

\bibitem[{{Couch} {et~al.}(2010){Couch}, {Pooley}, {Wheeler}, \&
  {Milosavljevic}}]{Couch+2010}
{Couch}, S.~M., {Pooley}, D., {Wheeler}, J.~C., \& {Milosavljevic}, M. 2010,
  ArXiv e-prints, 1007.3693

\bibitem[{{Galama} {et~al.}(1998){Galama}, {Vreeswijk}, {van Paradijs},
  {Kouveliotou}, {Augusteijn}, {B{\"o}hnhardt}, {Brewer}, {Doublier},
  {Gonzalez}, {Leibundgut}, {Lidman}, {Hainaut}, {Patat}, {Heise}, {in't Zand},
  {Hurley}, {Groot}, {Strom}, {Mazzali}, {Iwamoto}, {Nomoto}, {Umeda},
  {Nakamura}, {Young}, {Suzuki}, {Shigeyama}, {Koshut}, {Kippen}, {Robinson},
  {de Wildt}, {Wijers}, {Tanvir}, {Greiner}, {Pian}, {Palazzi}, {Frontera},
  {Masetti}, {Nicastro}, {Feroci}, {Costa}, {Piro}, {Peterson}, {Tinney},
  {Boyle}, {Cannon}, {Stathakis}, {Sadler}, {Begam}, \& {Ianna}}]{Galama+1998}
{Galama}, T.~J. {et~al.} 1998, \nat, 395, 670, arXiv:astro-ph/9806175

\bibitem[{{Granot} \& {Loeb}(2003)}]{GranotL2003}
{Granot}, J., \& {Loeb}, A. 2003, \apjl, 593, L81, arXiv:astro-ph/0305379

\bibitem[{{Johnston} {et~al.}(2008){Johnston}, {Taylor}, {Bailes}, {Bartel},
  {Baugh}, {Bietenholz}, {Blake}, {Braun}, {Brown}, {Chatterjee}, {Darling},
  {Deller}, {Dodson}, {Edwards}, {Ekers}, {Ellingsen}, {Feain}, {Gaensler},
  {Haverkorn}, {Hobbs}, {Hopkins}, {Jackson}, {James}, {Joncas}, {Kaspi},
  {Kilborn}, {Koribalski}, {Kothes}, {Landecker}, {Lenc}, {Lovell}, {Macquart},
  {Manchester}, {Matthews}, {McClure-Griffiths}, {Norris}, {Pen}, {Phillips},
  {Power}, {Protheroe}, {Sadler}, {Schmidt}, {Stairs}, {Staveley-Smith},
  {Stil}, {Tingay}, {Tzioumis}, {Walker}, {Wall}, \&
  {Wolleben}}]{Johnston+2008}
{Johnston}, S. {et~al.} 2008, Experimental Astronomy, 22, 151, 0810.5187

\bibitem[{{Kifonidis} {et~al.}(2006){Kifonidis}, {Plewa}, {Scheck}, {Janka}, \&
  {M{\"u}ller}}]{Kifonidis+2006}
{Kifonidis}, K., {Plewa}, T., {Scheck}, L., {Janka}, H., \& {M{\"u}ller}, E.
  2006, \aap, 453, 661, arXiv:astro-ph/0511369

\bibitem[{{Kulkarni} {et~al.}(1998){Kulkarni}, {Frail}, {Wieringa}, {Ekers},
  {Sadler}, {Wark}, {Higdon}, {Phinney}, \& {Bloom}}]{Kulkarni+1998}
{Kulkarni}, S.~R. {et~al.} 1998, \nat, 395, 663

\bibitem[{{Levesque} {et~al.}(2010){Levesque}, {Soderberg}, {Foley}, {Berger},
  {Kewley}, {Chakraborti}, {Ray}, {Torres}, {Challis}, {Kirshner}, {Barthelmy},
  {Bietenholz}, {Chandra}, {Chaplin}, {Chevalier}, {Chugai}, {Connaughton},
  {Copete}, {Fox}, {Fransson}, {Grindlay}, {Hamuy}, {Milne}, {Pignata},
  {Stritzinger}, \& {Wieringa}}]{Levesque+2010}
{Levesque}, E.~M. {et~al.} 2010, \apjl, 709, L26, 0908.2818

\bibitem[{{Li} \& {Chevalier}(1999)}]{LiC1999}
{Li}, Z., \& {Chevalier}, R.~A. 1999, \apj, 526, 716, arXiv:astro-ph/9903483

\bibitem[{{Malesani} {et~al.}(2004){Malesani}, {Tagliaferri}, {Chincarini},
  {Covino}, {Della Valle}, {Fugazza}, {Mazzali}, {Zerbi}, {D'Avanzo},
  {Kalogerakos}, {Simoncelli}, {Antonelli}, {Burderi}, {Campana}, {Cucchiara},
  {Fiore}, {Ghirlanda}, {Goldoni}, {G{\"o}tz}, {Mereghetti}, {Mirabel},
  {Romano}, {Stella}, {Minezaki}, {Yoshii}, \& {Nomoto}}]{Malesani+2004}
{Malesani}, D. {et~al.} 2004, \apjl, 609, L5, arXiv:astro-ph/0405449

\bibitem[{{Mart{\'{\i}}-Vidal} \& {Marcaide}(2008)}]{Marti-VidalM2008}
{Mart{\'{\i}}-Vidal}, I., \& {Marcaide}, J.~M. 2008, \aap, 480, 289, 0801.1272

\bibitem[{{Massi} \& {Aaron}(1999)}]{MassiA1999}
{Massi}, M., \& {Aaron}, S. 1999, \aaps, 136, 211

\bibitem[{{Mazzali} {et~al.}(2010){Mazzali}, {Maurer}, {Valenti}, {Kotak}, \&
  {Hunter}}]{Mazzali+2010}
{Mazzali}, P.~A., {Maurer}, I., {Valenti}, S., {Kotak}, R., \& {Hunter}, D.
  2010, \mnras, 1078, 1006.4259

\bibitem[{{Paragi} {et~al.}(2005){Paragi}, {Garrett}, {Paczy{\'n}ski},
  {Kouveliotou}, {Szomoru}, {Reynolds}, {Parsley}, \& {Ghosh}}]{Paragi+2005}
{Paragi}, Z., {Garrett}, M.~A., {Paczy{\'n}ski}, B., {Kouveliotou}, C.,
  {Szomoru}, A., {Reynolds}, C., {Parsley}, S.~M., \& {Ghosh}, T. 2005, Memorie
  della Societa Astronomica Italiana, 76, 570, arXiv:astro-ph/0505468

\bibitem[{{Paragi} {et~al.}(2010){Paragi}, {Taylor}, {Kouveliotou}, {Granot},
  {Ramirez-Ruiz}, {Bietenholz}, {van der Horst}, {Pidopryhora}, {van
  Langevelde}, {Garrett}, {Szomoru}, {Argo}, {Bourke}, \&
  {Paczy{\'n}ski}}]{SN2007gr_Nature}
{Paragi}, Z. {et~al.} 2010, \nat, 463, 516, 1001.5060

\bibitem[{{Paragi} {et~al.}(2008){Paragi}, {van der Horst}, {Kouveliotou},
  {Garrett}, {Wijers}, {Granot}, {Ramirez-Ruiz}, \& {Strom}}]{Paragi+2008}
{Paragi}, Z., {van der Horst}, A., {Kouveliotou}, C., {Garrett}, M., {Wijers},
  R.~A.~M.~J., {Granot}, J., {Ramirez-Ruiz}, E., \& {Strom}, R. 2008, in The
  role of VLBI in the Golden Age for Radio Astronomy

\bibitem[{{Paturel} {et~al.}(2003){Paturel}, {Petit}, {Prugniel}, {Theureau},
  {Rousseau}, {Brouty}, {Dubois}, \& {Cambr{\'e}sy}}]{Paturel+2003}
{Paturel}, G., {Petit}, C., {Prugniel}, P., {Theureau}, G., {Rousseau}, J.,
  {Brouty}, M., {Dubois}, P., \& {Cambr{\'e}sy}, L. 2003, \aap, 412, 45

\bibitem[{{Petrov} {et~al.}(2006){Petrov}, {Kovalev}, {Fomalont}, \&
  {Gordon}}]{Petrov+2006}
{Petrov}, L., {Kovalev}, Y.~Y., {Fomalont}, E.~B., \& {Gordon}, D. 2006, \aj,
  131, 1872, arXiv:astro-ph/0508506

\bibitem[{{Pian} {et~al.}(2006){Pian}, {Mazzali}, {Masetti}, {Ferrero},
  {Klose}, {Palazzi}, {Ramirez-Ruiz}, {Woosley}, {Kouveliotou}, {Deng},
  {Filippenko}, {Foley}, {Fynbo}, {Kann}, {Li}, {Hjorth}, {Nomoto}, {Patat},
  {Sauer}, {Sollerman}, {Vreeswijk}, {Guenther}, {Levan}, {O'Brien}, {Tanvir},
  {Wijers}, {Dumas}, {Hainaut}, {Wong}, {Baade}, {Wang}, {Amati}, {Cappellaro},
  {Castro-Tirado}, {Ellison}, {Frontera}, {Fruchter}, {Greiner}, {Kawabata},
  {Ledoux}, {Maeda}, {M{\o}ller}, {Nicastro}, {Rol}, \& {Starling}}]{Pian+2006}
{Pian}, E. {et~al.} 2006, \nat, 442, 1011, arXiv:astro-ph/0603530

\bibitem[{{Pignata} {et~al.}(2009{\natexlab{a}}){Pignata}, {Maza}, {Antezana},
  {Cartier}, {Folatelli}, {Forster}, {Gonzalez}, {Gonzalez}, {Hamuy}, {Iturra},
  {Lopez}, {Silva}, {Conuel}, {Crain}, {Foster}, {Ivarsen}, {Lacluyze},
  {Nysewander}, \& {Reichart}}]{Pignata+2009b}
{Pignata}, G. {et~al.} 2009{\natexlab{a}}, in American Institute of Physics
  Conference Series, Vol. 1111, American Institute of Physics Conference
  Series, ed. {G.~Giobbi, A.~Tornambe, G.~Raimondo, M.~Limongi,
  L.~A.~Antonelli, N.~Menci, \& E.~Brocato}, 551--554

\bibitem[{{Pignata} {et~al.}(2009{\natexlab{b}}){Pignata}, {Maza}, {Hamuy},
  {Antezana}, {Gonzalez}, {Gonzalez}, {Lopez}, {Silva}, {Folatelli}, {Iturra},
  {Cartier}, {Forster}, {Marchi}, {Conuel}, {Reichart}, {Ivarsen}, {Crain},
  {Foster}, {Nysewander}, \& {Lacluyze}}]{Pignata+2009a}
{Pignata}, G. {et~al.} 2009{\natexlab{b}}, Central Bureau Electronic Telegrams,
  1731, 1

\bibitem[{{Schinzel} {et~al.}(2009){Schinzel}, {Taylor}, {Stockdale}, {Granot},
  \& {Ramirez-Ruiz}}]{Schinzel+2008}
{Schinzel}, F.~K., {Taylor}, G.~B., {Stockdale}, C.~J., {Granot}, J., \&
  {Ramirez-Ruiz}, E. 2009, \apj, 691, 1380, 0810.1478

\bibitem[{{Soderberg} {et~al.}(2010{\natexlab{a}}){Soderberg}, {Brunthaler},
  {Nakar}, {Chevalier}, \& {Bietenholz}}]{SN2007gr-Soderberg}
{Soderberg}, A.~M., {Brunthaler}, A., {Nakar}, E., {Chevalier}, R.~A., \&
  {Bietenholz}, M.~F. 2010{\natexlab{a}}, ArXiv e-prints, 1005.1932

\bibitem[{{Soderberg} {et~al.}(2010{\natexlab{b}}){Soderberg}, {Chakraborti},
  {Pignata}, {Chevalier}, {Chandra}, {Ray}, {Wieringa}, {Copete}, {Chaplin},
  {Connaughton}, {Barthelmy}, {Bietenholz}, {Chugai}, {Stritzinger}, {Hamuy},
  {Fransson}, {Fox}, {Levesque}, {Grindlay}, {Challis}, {Foley}, {Kirshner},
  {Milne}, \& {Torres}}]{SN2009bb_Nature}
{Soderberg}, A.~M. {et~al.} 2010{\natexlab{b}}, \nat, 463, 513, 0908.2817

\bibitem[{{Soderberg} {et~al.}(2006{\natexlab{a}}){Soderberg}, {Kulkarni},
  {Nakar}, {Berger}, {Cameron}, {Fox}, {Frail}, {Gal-Yam}, {Sari}, {Cenko},
  {Kasliwal}, {Chevalier}, {Piran}, {Price}, {Schmidt}, {Pooley}, {Moon},
  {Penprase}, {Ofek}, {Rau}, {Gehrels}, {Nousek}, {Burrows}, {Persson}, \&
  {McCarthy}}]{Soderberg+2006c}
------. 2006{\natexlab{a}}, \nat, 442, 1014, arXiv:astro-ph/0604389

\bibitem[{{Soderberg} {et~al.}(2006{\natexlab{b}}){Soderberg}, {Nakar},
  {Berger}, \& {Kulkarni}}]{Soderberg+2006b}
{Soderberg}, A.~M., {Nakar}, E., {Berger}, E., \& {Kulkarni}, S.~R.
  2006{\natexlab{b}}, \apj, 638, 930, arXiv:astro-ph/0507147

\bibitem[{{Stanek} {et~al.}(2003){Stanek}, {Matheson}, {Garnavich}, {Martini},
  {Berlind}, {Caldwell}, {Challis}, {Brown}, {Schild}, {Krisciunas}, {Calkins},
  {Lee}, {Hathi}, {Jansen}, {Windhorst}, {Echevarria}, {Eisenstein}, {Pindor},
  {Olszewski}, {Harding}, {Holland}, \& {Bersier}}]{Stanek+2003}
{Stanek}, K.~Z. {et~al.} 2003, \apjl, 591, L17, arXiv:astro-ph/0304173

\bibitem[{{Starling} {et~al.}(2010){Starling}, {Wiersema}, {Levan}, {Sakamoto},
  {Bersier}, {Goldoni}, {Oates}, {Rowlinson}, {Campana}, {Sollerman}, {Tanvir},
  {Malesani}, {Fynbo}, {Covino}, {D'Avanzo}, {O'Brien}, {Page}, {Osborne},
  {Vergani}, {Barthelmy}, {Burrows}, {Cano}, {Curran}, {De Pasquale}, {D'Elia},
  {Evans}, {Flores}, {Fruchter}, {Garnavich}, {Gehrels}, {Gorosabel}, {Hjorth},
  {Holland}, {van der Horst}, {Jakobsson}, {Kamble}, {Kuin}, {Kaper},
  {Mazzali}, {Nugent}, {Pian}, {Thoene}, \& {Woosley}}]{Starling+2010}
{Starling}, R.~L.~C. {et~al.} 2010, ArXiv e-prints, 1004.2919

\bibitem[{{Stritzinger} {et~al.}(2009){Stritzinger}, {Philips}, {Morrell},
  {Salgado}, \& {Folatelli}}]{Stritzinger+2009}
{Stritzinger}, M., {Philips}, M.~M., {Morrell}, N., {Salgado}, F., \&
  {Folatelli}, G. 2009, Central Bureau Electronic Telegrams, 1751, 1

\bibitem[{{Taylor} {et~al.}(2004){Taylor}, {Frail}, {Berger}, \&
  {Kulkarni}}]{Taylor+2004}
{Taylor}, G.~B., {Frail}, D.~A., {Berger}, E., \& {Kulkarni}, S.~R. 2004,
  \apjl, 609, L1

\bibitem[{{Ulvestad} {et~al.}(2010){Ulvestad}, {Romney}, {Brisken}, {Deller},
  {Walker}, \& {Durand}}]{Ulvestad+2010}
{Ulvestad}, J.~S., {Romney}, J.~D., {Brisken}, W.~F., {Deller}, A.~T.,
  {Walker}, R.~C., \& {Durand}, S.~J. 2010, in Bulletin of the American
  Astronomical Society, Vol.~41, Bulletin of the American Astronomical Society,
  407

\end{thebibliography}

\clearpage

\end{document}